# A convenient sealing method using boron nitride capping for reactive reactions

Boqin Song[1], Tianping Ying[1,*]

1. Beijing National Laboratory for Condensed Matter Physics, Institute of Physics, Chinese Academy of Sciences, Beijing 100190, China

*ying@iphy.ac.cn

## Abstract

**While quartz ($SiO_2$) ampoule sealing is commonly used in laboratories to prevent sample oxidation during synthesis, its application is limited for reactions involving highly reactive elements such as alkali, alkali-earth and rare-earth metals. These elements can react with $SiO_2$ at elevated temperatures, causing compositional loss, tube failure, and experimental inconsistencies. Here, we introduce an inexpensive boron nitride (BN) cap sealing technique. This approach is readily adaptable to centrifugal separation and flux transport growth, and yields superior sample quality. We demonstrate its efficacy by growing $KFe_2As_2$ and $CsCr_6Sb_6$ single crystals, the former exhibiting record-high quality, with a residual resistivity ratio (RRR) exceeding 2500, and the latter achieving significantly larger crystal dimension than other methods. This accessible and economical method promises to accelerate the discovery of novel materials containing reactive elements.**

## 1. Introduction

In the synthesis of air-sensitive materials, effective oxygen exclusion is essential. Sealing samples within $SiO_2$ ampoules is a widely adopted method to achieve this isolation[1-4]. However, this approach becomes inadequate for compounds containing highly reactive elements such as alkali[5-7], alkaline earth[7,8], or rare-earth metals[9,10]. At elevated temperatures, their high vapor pressure leads to inevitable composition deviation, while their chemical reactivity with silica further risks ampoule rupture and subsequent experimental failure. Carbon coating layer was found to prevent the silica ampoule but it also fails during long time heating. Alternative methods employing sealed Ta or Nb crucibles mitigate these issues but require costly metals and arc melting apparatus[11-15], with additional risks of air exposure during transfer from gloveboxes. Another approach involves sealing $Al_2O_3$ crucibles in a floating-zone furnace, which demands specialized and expensive equipment. These limitations significantly hinder the exploration of new materials.

Here, we present a low-cost, user-friendly sealing strategy that employs BN caps fitted onto conventional $Al_2O_3$ or similar crucibles. This design retains the advantages of inert-atmosphere sealing while avoiding the expense and refractory metal tubes. BN can be readily shaped into custom sieves for post centrifugal separation or fitted into tubes for flux-based transport growth. To showcase the efficacy and broad applicability of our method, we report the growth of high-quality single crystals of the superconductor $KFe_2As_2$, which achieves exceptional electronic quality exceeding that of Fe-sealed samples, and the Kagome Kondo lattice system $CsCr_6Sb_6$ yielding sizable crystals previously inaccessible via $SiO_2$ sealing. The design schematics of a user-friendly mechanical fixture is provided for convenient assembly of the BN sealing cap. We anticipate this practical and economical sealing technique will stimulate broader research into novel compounds incorporating reactive constituents.

## 2. Results

### 2.1 Excellent sealing performance of BN caps

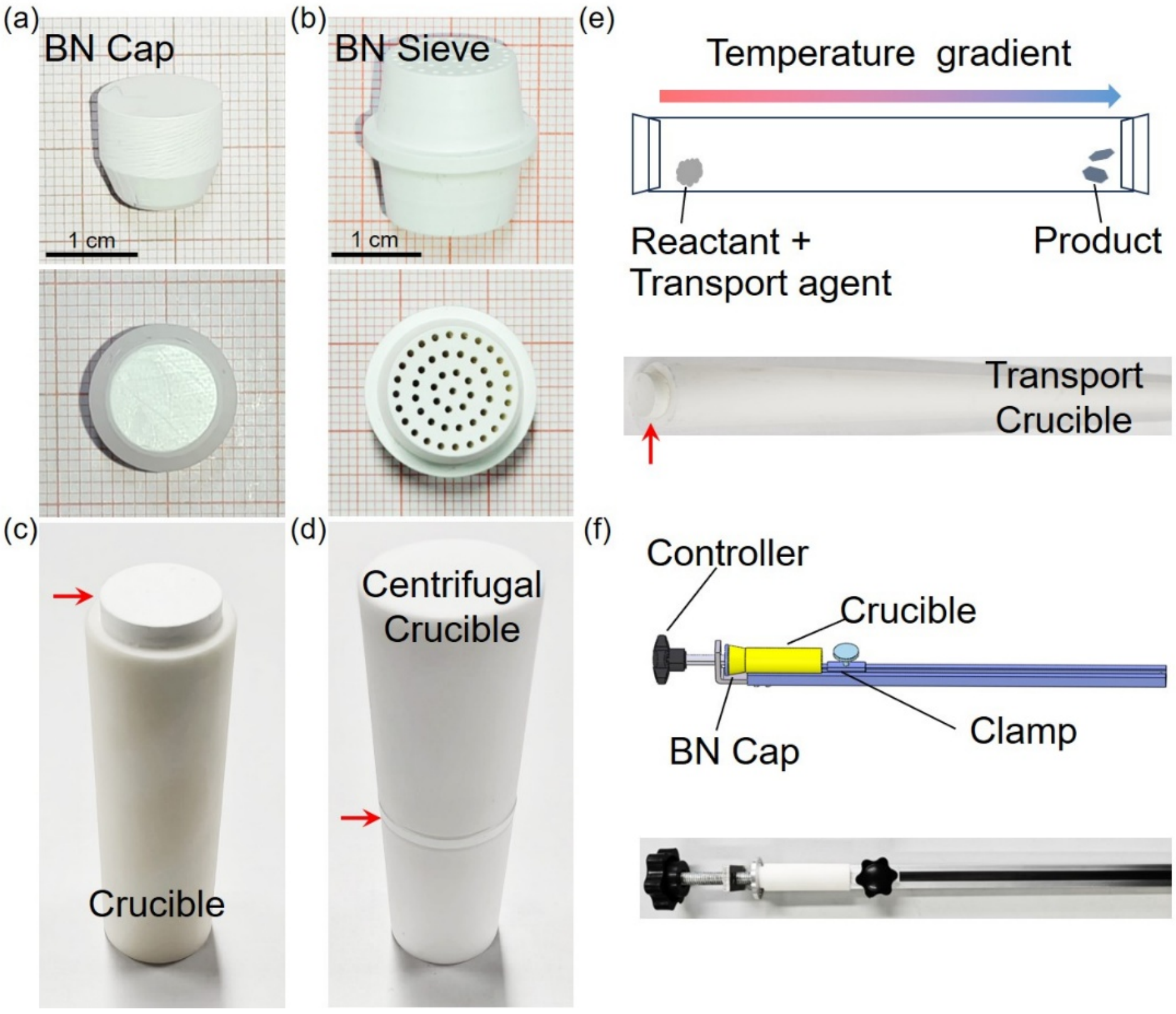

Figure 1. (a)-(b) BN cap and centrifugal sieve machined using laboratory equipment. Upper and lower panels show side and top views, respectively. (c)-(d) BN-sealed $Al_2O_3$ crucible and centrifugal crucible. Red arrows indicate the BN caps. (e) Schematic and optical image of the transport crucible, with red arrows denoting the BN cap. (f) Custom-made assembly tool to press-fit the BN caps onto the crucibles. Mechanical drawings of the assembly tool are provided in the Supplementary Information.

Our sealing strategy employs hexagonal BN (h-BN) as an inert and economical sealing cap for conventional crucibles such as $Al_2O_3$. Commercially available BN ceramic rods are inherently soft, making them exceptionally suitable for shaping into custom configurations even with limited laboratory equipment. Figure 1a, b shows a machined BN crucible cap and centrifugal sieve fabricated using a lathe, which can seal either a conventional crucible or the transport crucible depicted in Figure 1c, e. For the specific $Al_2O_3$ crucible used in this study with an inner diameter of 7 mm, we designed the BN cap as a truncated cone with a bottom diameter of 6.8 mm and a top diameter of 7.3 mm, allowing for easy pre-assembly. The assembly is then mounted into a custom-made assembly tool and secured, as shown in Figure. 1f, with pressure applied by turning a threaded knob as demonstrated in the Supplementary Information video. Due to soft texture of BN, the cap deforms slightly when pressed against the rigid crucible, thereby achieving an effective and reliable seal.

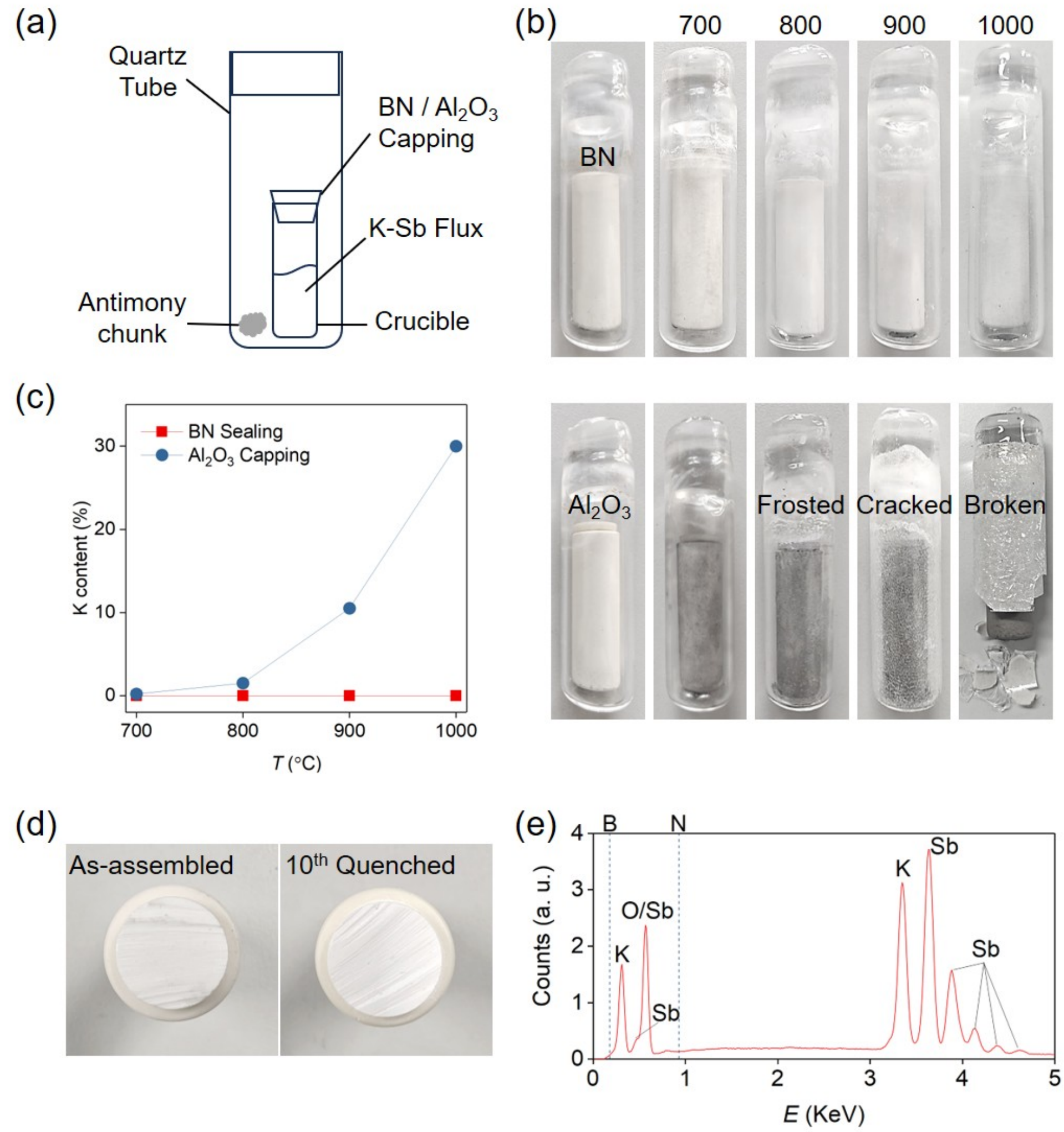

Figure 2. (a) Schematic illustration of BN seal in preventing alkali metal volatilization and tube corrosion. BN/$Al_2O_3$-capped crucibles containing K-Sb flux are sealed in quartz tubes together with Sb chunks to trap evaporated potassium vapor. (b) Controlled experiments comparing BN and $Al_2O_3$ capping at a series of temperatures. Tubes sealing with $Al_2O_3$-capped crucibles are attacked more violently with increasing temperature, while tubes sealing with BN-capped crucibles remain intact. (c) K content absorbed by the Sb chunk outside the crucible, as illustrated in (a). (d) Optical images of the assembled BN cap before and after 10 thermal quench cycles from 1000 °C to room temperature. (e) EDS map of the K-Sb flux in the BN-capped crucible after heating at 1000 °C, no B or N impurity are detected.

We demonstrate the excellent sealing performance and robustness of the BN cap using K-Sb flux, comparing it against conventional $Al_2O_3$ capping. Figure 2a provides a schematic illustration of the experimental setup. Two sets of $Al_2O_3$ crucibles, one capped with BN and the other with $Al_2O_3$, were each loaded with 4 g of potassium ingot and antimony powder mixture in a 1:3 stoichiometry. After capping, the crucibles were sealed in evacuated quartz tubes (20 mm inner diameter) and placed in a muffle furnace for heat treatment. To quantitatively monitor potassium leakage, a 0.5 g Sb chunk was placed at the bottom of each tube alongside the capped crucible; the K content absorbed by the Sb chunk was subsequently measured using energy-dispersive X-ray spectroscopy (EDS). All tubes were heated to target temperatures ranging from 700-1000 °C, held for 10 hours, and then quenched in air.

Figure 2b shows photographs of the quenched tubes at different temperatures. Tubes containing $Al_2O_3$-capped crucibles exhibited severe contamination from leaked reactants, with quartz tube walls visibly corroded after quenching and eventually fractured at temperatures above 1000 °C. In contrast, tubes with BN-capped crucibles remained clean, and the quartz walls stayed intact after quenching. Figure 2c presents the K content measured in the Sb chunk for each tube (performed in Phenom ProX EDS). For $Al_2O_3$-capped crucibles, the K content increased drastically with temperature, while BN-capped crucibles remained the same even after heating at 1000 °C. We also use a carbon coated quartz tube for comparation, which is also broken, see Supporting Information. These results indicate excellent thermal stability of the BN seal[16].

The BN caps also demonstrated outstanding thermal shock resistance, enabling robust sealing even after repeated quenching over a wide temperature range. As shown in Figure 2d, a BN-capped $Al_2O_3$ crucible was subjected to 10 consecutive quench cycles from 1100 °C to water, with the cap remaining intact and free of visible cracks or damage. Finally, elemental analysis was performed on the K-Sb flux inside the BN-capped crucible after heating at 1000 °C. The EDS map in Figure 2e reveals no detectable B or N impurities, confirming the superior chemical stability of the BN cap at high temperatures, a critical requirement for crystal synthesis in the absence of interfering contaminants.

### 2.2 Alternative sealing strategy for enhanced crystallinity

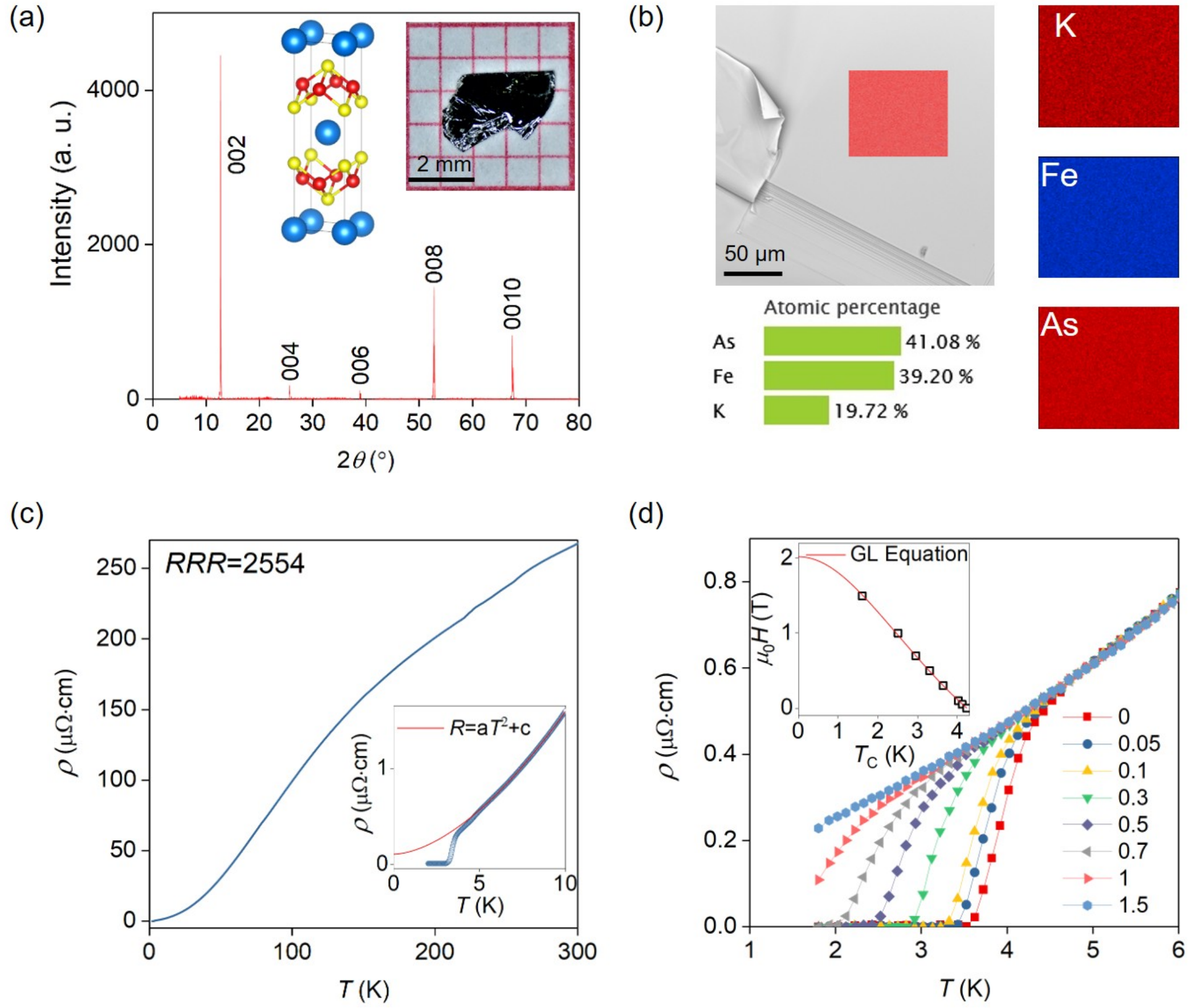

Figure 3. (a) XRD pattern showing the 00*l* reflections of $KFe_2As_2$ single crystal. Left panel of inset shows the crystal structure of $KFe_2As_2$. Right panel shows an optical image of the as-grown crystal. (b) EDS results from a freshly cleaved surface of the as-grown crystal, indicating stoichiometry of 1:2:2 within apparatus uncertainty. (c) Temperature-dependent resistivity of the $KFe_2As_2$ sample. A record-high RRR of 2554 can be extracted by fitting $\rho_0$ using the Fermi liquid model. (d) Resistivity under magnetic field. Inset shows the fit of the upper critical field using the Ginzburg-Landau equation.

The selection of an appropriate flux is critical in single crystal growth, as exemplified by the synthesis of $KFe_2As_2$ single crystals. $KFe_2As_2$ is the most heavily hole-doped member of the iron-based superconductors and exhibits properties distinct from its parent compound $BaFe_2As_2$ and optimally doped $Ba_{1-x}K_xFe_2As_2$, notably the absence of electronic nematicity driven by spin fluctuations[17]. However, this interpretation remains debated, as recent studies have suggested the possible presence of nematic order in this material[18-21], making high-quality, sizable single crystals essential to resolve this controversy. Although polycrystalline $KFe_2As_2$ can be readily synthesized, early attempts to grow single crystals using tin flux were frequently plagued by Sn contamination[22], while efforts using Fe-As self-flux encountered difficulties because the vapor pressure of K rises substantially at the melting point of the flux (1030 °C), leading to attack on the quartz tube and preventing stable, long-duration crystal growth[23], which resulted in limited crystal size and poor sample quality. While the use of K-As self-flux lowers the growth temperature to 900 °C and improves crystal quality, the large quantity of K required in this approach still necessitates a specialized stainless-steel container[24].

We demonstrate the effectiveness of our BN sealing method by synthesizing $KFe_2As_2$ single crystal using a highly K-rich flux. The stating materials are K ingot, Fe powder and As chunk without precursor preparation. Typically, 4 g of a mixture with a nominal stoichiometry of K:Fe:As=6:1:7 was placed in an $Al_2O_3$ crucible and sealed with a BN cap inside a glove box. The assembly was then flame-sealed in an evacuated quartz tube and placed in a furnace. The temperature was ramped to 900 °C over 10 hours, held for 24 hours and finally cooled at a rate of 1.5 °C/h. After growth, the flux was removed using ethanol. The typical size of the yielding crystals is 5×5×0.4 $mm^3$. As shown in the inset of Figure 3a, the grown crystal exhibits a plate-like morphology with flat, shiny surfaces that can be easily cleaved. An orthorhombic-geometry is recognized at the cleaved surface, indicating the exposure of the $00l$ plane. Figure 3a shows the X-ray diffraction pattern with sharp $00l$ reflections (collected in Rigaku Smart Lab diffractometer with Cu $K_\alpha$ radiation). The indexed lattice parameter $c$=13.88 Å is well consistent with previous studies. Figure 3b shows a scan electron microscopy (SEM) image of a cleaved surface, revealing clear layered feature. EDS confirms an elemental ratio consistent with the $KFe_2As_2$ stoichiometry within apparatus uncertainty.

Figure 3c displays the temperature dependent resistivity of a bulk $KFe_2As_2$ sample, which exhibits metallic behavior across the entire temperature range (measured in QD PPMS equipped with SR 830 Lock-in Amplifier). The resistivity decreases gradually around 3.9 K and reaches zero at $T_c$=3.2 K, which is superior to those reported in previous studies. Fitting the low-temperature resistivity using the Fermi liquid dependence $\rho=AT^2+\rho_0$ yields a residual resistivity $\rho_0$=0.1 μΩ·cm, which corresponds to a record-high RRR of 2554. This exceptionally high RRR indicates the as-grown single crystal is extremely clean and almost free from elastic impurity scattering. To compare with other methods, we calculate $\rho_{300\ K}/\rho_{T_c}$ which gives 725 for our single crystal, substantially higher than the value of 87 reported for the growth by the FeAs flux method[25]. Figure 3d shows the temperature dependent resistivity under various magnetic field. We define $T_c$ as the temperature at which the resistivity drops to 90% of the normal state and plot the resulting data in the inset. The Ginzburg-Landau (GL) equation is used to fit the upper critical field $H_{c2}$, yielding an approximate value of 2 T. The coherent length can be estimated from $H_{c2}$ using the relation $H_{c2}(0) = \phi_0/(2\pi\xi_0^2)$, where $\phi_0$ is the magnetic flux quantum and $\xi_0$ is the coherence length at zero temperature. This gives $\xi_0$=17 nm, confirming that our samples are in the clean limit.

## 2.3 Speed up exploration of novel materials

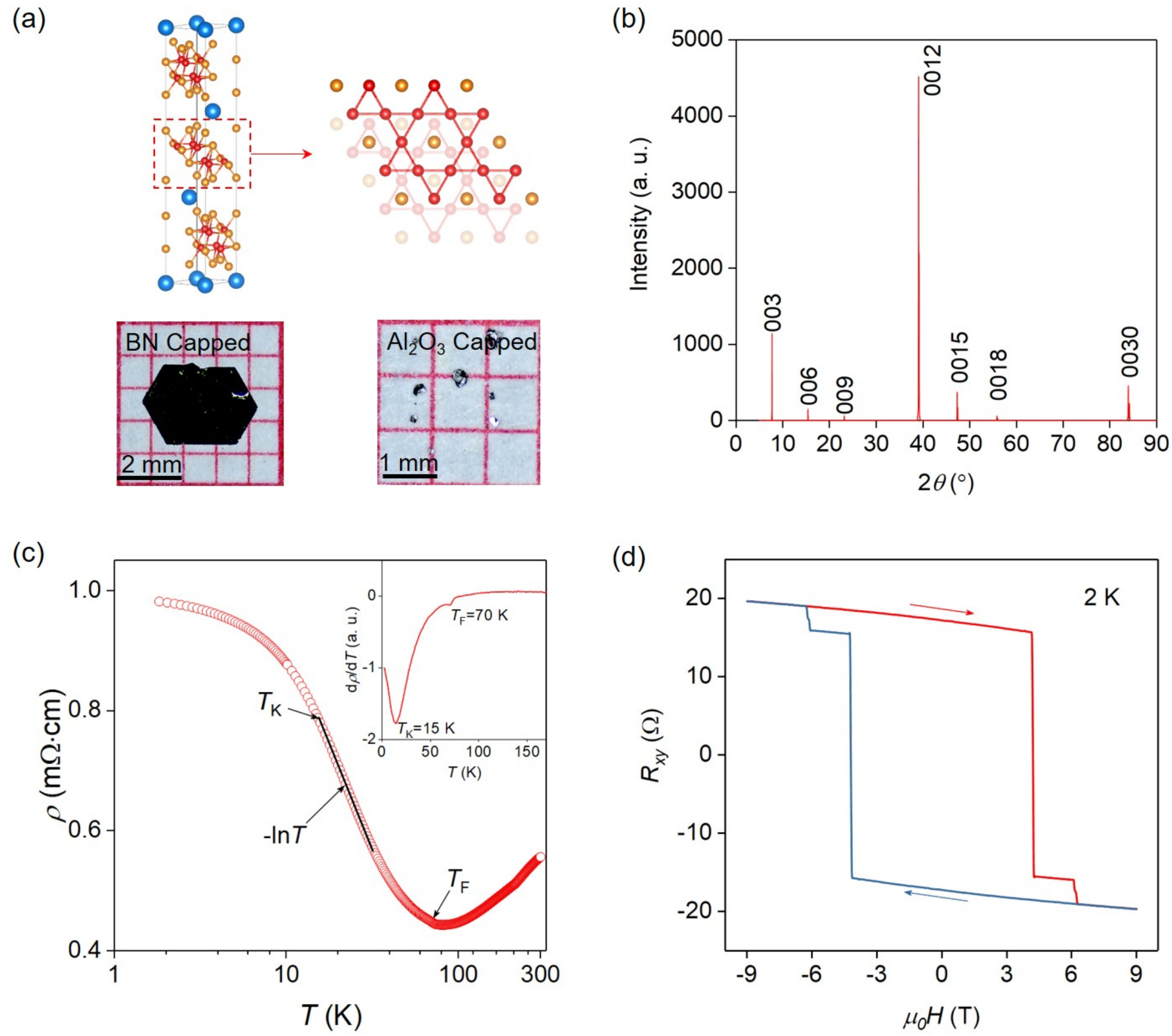

Figure 4. (a) Upper panels are the crystal structure $CsCr_6Sb_6$. Red dashed box is a double Kagome layer sandwich bay Cs layers, which consists two Kagome lattice of Cr. Lower panels show the Comparison of $CsCr_6Sb_6$ crystals grown with and without BN capping. (b) XRD pattern of the as-grown $CsCr_6Sb_6$ crystal, displaying a series of 00*l* reflections. (c) Temperature-dependent resistivity of $CsCr_6Sb_6$. Kondo insulating behavior is evidenced by the -ln$T$ dependence (black line). The Kondo temperature $T_K$ and the temperature associated with frustrated magnetic interactions $T_F$ are clearly resolved in the d$\rho$/d$T$ plot shown in the inset. (d) Anomalous Hall resistance $R_{xy}$ of a 3-layer sample exhibits an extremely sharp jump at the spin-flipping field.

Having demonstrated the effectiveness of BN caps in sealing alkali-rich fluxes for $KFe_2As_2$ growth, we find their utility extends to creating superstoichiometric environments that facilitate non-thermodynamic reactions. This capability is particularly valuable when working with oversaturated, reactive elemental species. Compared to costly and complex methods like Ta/Nb or arc-sealed $Al_2O_3$ tubes, our BN sealing method cost less than 10% as much and can be assembled in a standard glovebox. This simple, economical approach holds significant promise for accelerating new materials discovery in conventional laboratory settings.

We demonstrate the capability of our BN sealing method through comparative synthesis of our recently discovered Kagome Kondo lattice material, $CsCr_6Sb_6$[26]. While recent years have witnessed a surge of interest in Kagome metals such as the $CsV_3Sb_5$ family[27-32], attempts at chemical substitution have proven largely unsuccessful, with only Ti and Cr showing any appreciable incorporation[33,34]. The recent discovery of $CsCr_3Sb_5$ using Ta tube sealing further intensified this interest, but these attempts always suffer from extremely low yields and unpredict reproducibility, severely limiting the availability of high-quality samples for broader investigation.

Polycrystalline $CsCr_6Sb_6$ cannot be obtained through stoichiometric reactions, therefore, we employed a far exceeding amount of Cs to create an oversaturated Cs environment, as discussed above. Cs metal, Cr powder, and Sb chunks were weighed in a molar ratio of 10:3:30 and loaded into an alumina crucible. BN capping was applied, followed by quartz tube sealing in vacuum. The tubes were heated to 1000 ℃ in a furnace and held for 24 hours. Subsequently, the furnace was rapidly cooled to 800 ℃ over 2 hours, then slowly cooled to 650 ℃ at a rate of 1 ℃/h. Finally, the tube was quenched in air and carefully tilted. By decanting the molten flux, thicker crystals (~20 μm in thickness) settled at the bottom of the crucible became exposed on the flux surface. These crystals were collected by immersing the mixture in water. The left lower panel of Figure 4a shows a flat plate-like crystal exceeding 3 mm in the ab-plane. The centrifugal sieve can also be employed to simplify sample collection. We also attempted to grow $CsCr_6Sb_6$ using conventional $Al_2O_3$ crucible caps or sieves. In these cases, the yields were extremely low, and the resulting crystals were too small for most measurements, as shown in the right panel of Figure 4a. A possible explanation is that quartz is highly reactive with Cs at elevated temperatures, effectively depleting the oversaturated Cs environment necessary for the successful growth of $CsCr_6Sb_6$.

Figure 4b shows the XRD patterns of $00l$ reflections from which the indexed lattice parameter c is found to be consistent with previous study. The temperature-dependent resistivity of a bulk sample is shown in Figure 4c. A Kondo insulating behavior is evident from the logarithmic divergence at low temperature. The Kondo temperature $T_K$ can be defined as the deviation of the -ln$T$ behavior of the resistivity at 15 K, while short-range magnetic interactions give rise to a small kink of resistivity at the frustrated magnetic temperature $T_F$=70 K, and both features are captured from the temperature derivative of resistivity $d\rho/dT$ as shown in the inset. $T_K$ and $T_F$ are lightly smaller than the reported values. We also fabricate a thin flake device with thickness of 3 layers, in which a A-type antiferromagnetism emerges from the Kondo breakdown. Figure 4d shows $R_{xy}$ of the 3 L sample at various temperature. Compared with previous study, we observe sharper and more symmetric spin flip upon sweeping the filed, indicating well-defined magnetic order and reduced exchange bias. The absence of additional steps comparing to the previous results suggests a homogeneous magnetic domain structure[32].

Thus far, superconductivity, the Kondo effect, and some strongly correlated phenomena have been experimentally realized in Cr-, V-, and Ti-based transition-metal Kagome lattices[26,31-34]. Theoretically, dozens of other possible Kagome materials incorporating other transition metals have been predicted to be energetically stable and await experimental synthesis[35]. Our method paves the way for exploring reactive-element-containing compounds at an affordable cost, in Kagome materials and beyond.

## 3. Conclusions

In summary, we have developed a cost-effective, robust, and versatile BN cap sealing method for synthesizing materials containing highly reactive elements. This technique overcomes the limitations of conventional $SiO_2$ ampoule sealing, which fails with alkali, alkaline-earth, and rare-earth metals due to reactive vapor attack, while avoiding expensive refractory metals or complex equipment. The simplicity, low cost, and adaptability of our BN caps make them a powerful tool for materials exploration. We anticipate that this "poor man's sealing method" will lower the barrier to synthesizing reactive-element compounds, accelerating the discovery of novel quantum materials.

## Supporting Information

Results of a control experiment using carbon coated tube and a video of assembling of the BN caps are included in Supporting Information.

## Note

Although a domestic patent has been applied for regarding the BN sealing method described herein [patent No. 202610267026.5], we grant permission for free use of this method, provided that this paper is appropriately cited. Additionally, we have designed an assembly tool to facilitate the sealing process. The corresponding mechanical drawing is available in the Supporting Information. Interested readers are welcome to contact us to request it free of charge (shipping not included).

## Acknowledgements

This work is financially supported by the National Key Research and Development Program of China (No. 2021YFA1401800), National Natural Science Foundation of China (No. 52522201).

## References

1. Good, W. D.; Scott, D. W.; Waddington, G. Combustion calorimetry of organic fluorine compounds by a rotating-bomb method. *J. Phys. Chem.* **1956**, 60, 1080–1089.
2. P. Rudolph, F.M. Kiessling The horizontal Bridgman method. *Cryst. Res. Technol.* **1988,** 23, 1207-1224.
3. Kirkpatrick, R. Crystal growth from the melt: a review. *Am. Mineral.* **1975**, 60, 798–814.
4. Hulliger, J. Chemistry and crystal growth. *Angew. Chem., Int. Ed.* **1994**, 33, 143–162,
5. Guo, J. G.; Jin, S. F.; Wang, G.; Wang, S. C.; Zhu, K. X.; Zhou, T. T.; He, M.; Chen, X. L. Phys. Rev. B: Condens. Matter Mater. Phys. **2010**, 82, 180520 DOI: 10.1103/PhysRevB.82.180520.
6. Ying, T. P.; Chen, X. L.; Wang, G.; Jin, S. F.; Zhou, T. T.; Lai, X. F.; Zhang, H.; Wang, W. Y. Sci. Rep. **2012**, 2, 426 DOI: 10.1038/srep00426
7. Ying, T. P.; Chen, X. L.; Wang, G.; Jin, S. F.; Lai, X. F.; Zhou, T. T.; Zhang, H.; Shen, S. J.; Wang, W. Y. J. Am. Chem. Soc. **2013**, 135, 2951– 2954 DOI: 10.1021/ja312705x.
8. Takahashi, H.; Sugimoto, A.; Nambu, Y.; Yamauchi, T.; Hirata, Y.; Kawakami, T.; Avdeev, M.; Matsubayashi, K.; Du, F.; Kawashima, C.; Soeda, H.; Nakano, S.; Uwatoko, Y.; Ueda, Y.; Sato, T. J.; Ohgushi, K. Pressure-induced superconductivity in the iron-based ladder material $BaFe_2S_3$. *Nat.Mater.* **2015**, 14, 1008−1012.
9. Zhou,B. B.; Misra, S. E.; Neto, H. da Silva; Aynajian, P.; Baumbach, R. E.; Thompson, J. D.; Bauer, E. D.; Yazdani, A. Visualizing nodal heavy fermion superconductivity in $CeCoIn_5$. *Nat. Phys.* **2013**, 9, 474.
10. Shen, B.; Zhang, Y.; Komijani, Y.; Nicklas, M.; Borth, R.; Wang, A.; Chen, Y.; Nie,

Z.; Li, R.; Lu, X.; Lee, H.; Smidman, M.; Steglich, F.; Coleman, P.; Yuan, H. Strange-metal behaviour in a pure ferromagnetic Kondo lattice. *Nature* **2020**, *579*, 51– 55, DOI: 10.1038/s41586-020-2052-z.

11. Kliemt, K. et al. Crystal growth of materials with the $ThCr_2Si_2$ structure type. *Cryst. Res. Technol.* **55**, 1900116 (2020).
12. Bao, J.-K.; Liu, J.-Y.; Ma, C.-W.; Meng, Z.-H.; Tang, Z.-T.; Sun, Y.-L.; Zhai, H.-F.; Jiang, H.; Bai, H.; Feng, C.-M. Superconductivity in Quasi-One-Dimensional $K_2Cr_3As_3$ with Significant Electron Correlations. *Phys. Rev. X* **2015**, *5*, 011013 DOI: 10.1103/PhysRevX.5.011013.
13. Yi Liu et al. Superconductivity under pressure in a chromium-based kagome metal. *Nature* **632**, 1032–1037 (2024).
14. Wang, Z.-C.; He, C.-Y.; Wu, S.-Q.; Tang, Z.-T.; Liu, Y.; Ablimit, A.; Feng, C.-M.; Cao, G.-H. Superconductivity in $KCa_2Fe_4As_4F_2$ with Separate Double $Fe_2As_2$ Layers. *J. Am. Chem. Soc.* **2016**, *138*, 7856–7859, DOI: 10.1021/jacs.6b04538.
15. Meng, K.; Zhang, X.; Song, B.; Li, B.; Kong, X.; Huang, S.; Yang, X.; Jin, X.; Wu, Y.; Nie, J. Layer-Dependent Superconductivity in Iron-Based Superconductors $Ca_2Fe_4As_4F_2$ and $CaKFe_4As_4$. *Nano Lett.* **2024**, *24*, 6821– 6827, DOI: 10.1021/acs.nanolett.4c01725.
16. Kostoglou, N.; Polychronopoulou, K.; Rebholz, C. Thermal and chemical stability of hexagonal boron nitride (h-BN) nanoplatelets. *Vacuum* **2015**, *112*, 42– 45, DOI: 10.1016/j.vacuum.2014.11.009.
17. Fernandes, R., Chubukov, A. & Schmalian, J. What drives nematic order in iron-based superconductors? *Nat. Phys.* **2014**, 10, 97–104.
18. Borisov, V., Fernandes, R. M. & Valentí, R. Evolution from $B_{2g}$ nematics to $B_{1g}$ nematics in heavily hole-doped iron-based superconductors. *Phys. Rev. Lett.* **2019**, 123, 146402.
19. Moroni, M. et al. Charge and nematic orders in $AFe_2As_2$ (A = Rb, Cs) superconductors. *Phys. Rev. B* **2019**, 99, 235147.
20. Liu, X., Tao, R., Ren, M. *et al.* Evidence of nematic order and nodal superconducting gap along [110] direction in $RbFe_2As_2$. *Nat. Commun.* **2019**, 10, 1039.
21. Yuta Mizukami, Ohei Tanaka, Kousuke Ishida, Asato Onishi, Yoichi Kageyama, Masaya Tsujii, Ryotaro Ohno, Noriaki Kimura, Takaya Mitsui, Shinji Kitao, Masayuki Kurokuzu, Makoto Seto, Shigeyuki Ishida, Akira Iyo, Hiroshi Eisaki, Kenichiro Hashimoto, Takasada Shibauchi, Thermodynamic signatures of diagonal nematicity in $RbFe_2As_2$ superconductor, *PNAS* **2025**, 4, pgaf060.

22. N. Ni, S. L. Bud'ko, A. Kreyssig, S. Nandi, G. E. Rustan, A. I. Goldman, S. Gupta, J. D. Corbett, A. Kracher, and P. C. Canfield Anisotropic thermodynamic and transport properties of single-crystalline $Ba_{1-x}K_xFe_2As_2$ (x=0 and 0.45). *Phys. Rev. B* **2008**, 78, 014507.
23. X. F. Wang, T. Wu, G. Wu, H. Chen, Y. L. Xie, J. J. Ying, Y. J. Yan, R. H. Liu, and X. H. Chen Anisotropy in the Electrical Resistivity and Susceptibility of Superconducting $BaFe_2As_2$ Single Crystals. *Phys. Rev. Lett.* **2009**, 102, 117005.
24. Kihou, K.; Saito, T.; Ishida, S.; Nakajima, M.; Tomioka, Y.; Fukazawa, H.; Kohori, Y.; Ito, T.; Uchida, S.; Iyo, A.; Lee, C.-H.; Eisaki, H. Single Crystal Growth and Characterization of the Iron-Based Superconductor $KFe_2As_2$ Synthesized by KAs Flux Method *J. Phys. Soc. Jpn.* **2010**, 79, 124713 DOI: 10.1143/JPSJ.79.124713.
25. J. K. Dong, S. Y. Zhou, T. Y. Guan, H. Zhang, Y. F. Dai, X. Qiu, X. F. Wang, Y. He, X. H. Chen, and S. Y. Li Quantum Criticality and Nodal Superconductivity in the FeAs-Based Superconductor $KFe_2As_2$. *Phys. Rev. Lett.* **2010**, 104,087005.
26. Song, B., Xie, Y., Li, WJ. *et al.* Realization of Kagome Kondo lattice. *Nat. Commun.* **2025**, 16, 5643.
27. Ye, L. D. et al. Massive Dirac fermions in a ferromagnetic kagome metal. *Nature* **2018**, 555, 638–642.
28. Yin, J. X. et al. Giant and anisotropic many-body spin–orbit tunability in a strongly correlated kagome magnet. *Nature* **2018**, 562, 91–95.
29. Morali, N. et al. Fermi-arc diversity on surface terminations of the magnetic Weyl semimetal $Co_3Sn_2S_2$. *Science* **2019**, 365, 1286–1291.
30. Liu, D. F. et al. Magnetic Weyl semimetal phase in a Kagomé crystal. *Science* **2019**, 365, 1282–1285.
31. Ortiz, B. R. *et al.* $CsV_3Sb_5$: a $z_2$ topological kagome metal with a superconducting ground state. *Phys*. *Rev*. *Lett*. **2020**, 125, 247002.
32. Song, B., Ying, T., Wu, X. *et al.* Anomalous enhancement of charge density wave in kagome superconductor $CsV_3Sb_5$ approaching the 2D limit. *Nat. Commun.* **2023**, 14, 2492.
33. Yang, H., Ye, Y., Zhao, Z. *et al.* Superconductivity and nematic order in a new titanium-based kagome metal $CsTi_3Bi_5$ without charge density wave order. *Nat. Commun.* **2024**, 15, 9626.
34. Yi Liu *et al.* Superconductivity under pressure in a chromium-based Kagome metal. *Nature* **2024**, 632, 1032–1037.
35. Cai G, Jiang Y, Zhou H, et al. Energy Landscape and Phase Competition of

CsV3Sb5, CsV6Sb6 and TbMn6Sn6-Type Kagome Materials. *Chin. Phys. Lett.* **2023**, 40, 117101.

**For Table of Contents Use Only**

# A convenient sealing method using boron nitride capping for reactive reactions

Boqin Song[1], Tianping Ying[1,*]

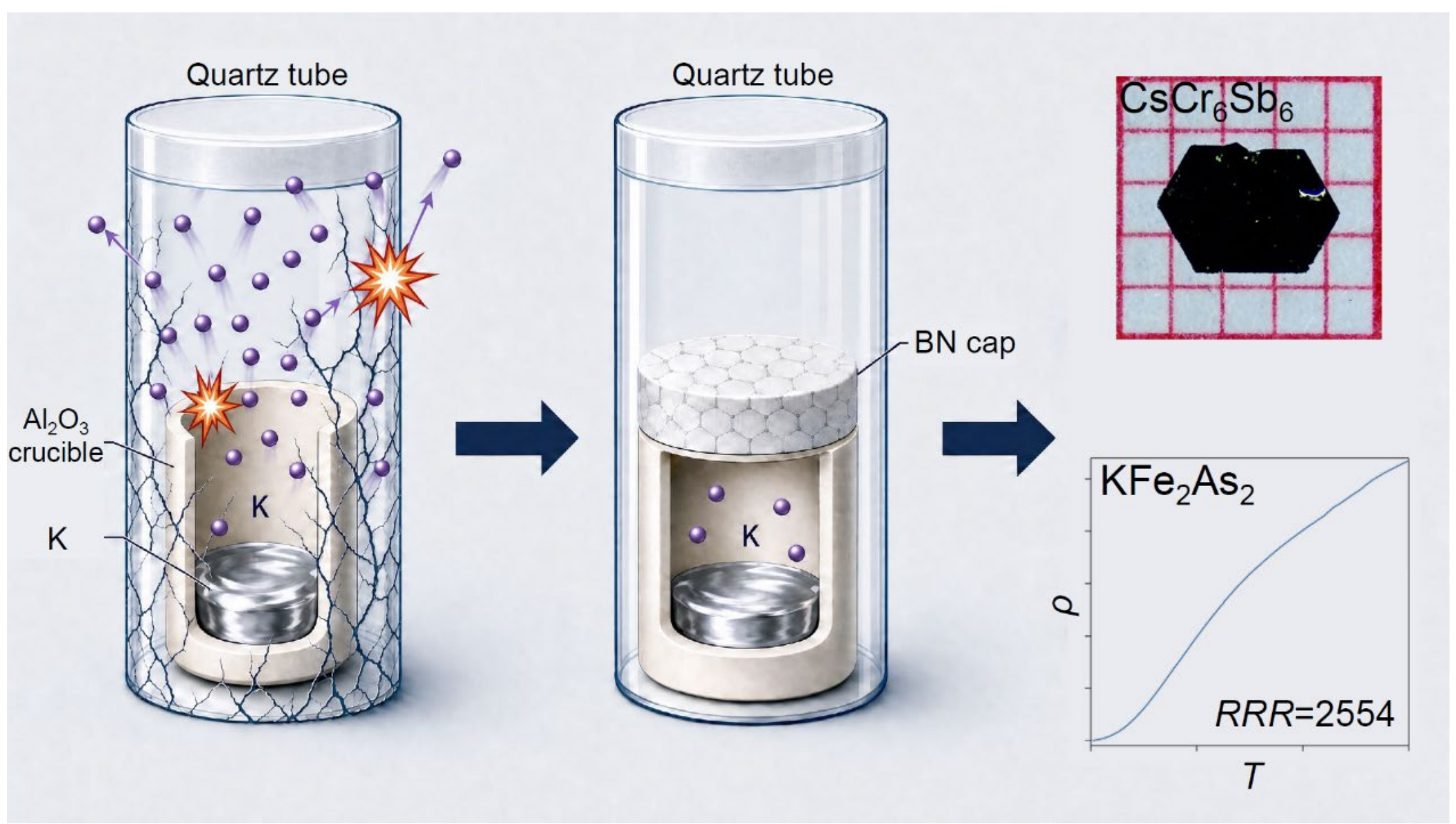